\documentclass[10pt,letterpaper]{article}
\usepackage{spconf,amsmath,amssymb,graphicx,booktabs}
\usepackage[T1]{fontenc}
\usepackage[utf8]{inputenc}
\usepackage{cite,url,capt-of,float}
\usepackage[hidelinks]{hyperref}
\newlength{\panelwidth}
\newcommand{\panel}[3]{
    \begin{minipage}[t]{\panelwidth}\centering\includegraphics[width=\linewidth]{#1}\\{\small\textbf{(#2)}
        #3}
\end{minipage}}
\title{Expert Spindles in the Wild: Enriched Supervision for a Dense Detector}
\name{
    Olivier Pallanca$^{\star}$,
    Alexis Bollengier$^{\dagger \star}$,
    Abel Diaz Berenguer$^{\dagger}$,
    Hichem Sahli$^{\dagger \ddagger}$,
    André Anjos$^{\star}$
}

\address{
    $^{\star}$ Idiap Research Institute, Martigny, Switzerland\\
    $^{\dagger}$ Department of Electronics and Informatics (ETRO), Vrije
    Universiteit Brussel (VUB), Brussels, Belgium\\
    $^{\ddagger}$ Interuniversity Microelectronics Centre (IMEC), Heverlee, Belgium
}
\begin{document}
\raggedbottom
\maketitle

\begin{abstract}
    Sleep spindles are a hallmark of healthy NREM sleep whose frequency,
    amplitude, density, duration and topography vary with brain maturation,
    age and disease. Their manual annotation is impractical and variable
    between scorers, yet the public references used to train automatic
    detectors cover these characteristics only partially: stage N2 only, a
    single central derivation, or short pre-selected segments. Measuring
    duration and density requires dense detectors that learn event boundaries
    from annotated onsets and offsets, and therefore inherit the protocol
    under which those boundaries were scored. We introduce Expert Spindles in the Wild (ESW), a stage-agnostic expert spindle annotation set covering 19 complete MASS-SS2 recordings ($\approx$150\,h of continuous EEG, 29,819 events). A sleep neurophysiologist annotated the recordings using the full EEG montage, blinded to sleep-stage labels and prior spindle annotations. On the 15 nights shared
    with historical scorers, ESW recovers 92.7\% of E1 and 73.7\% of E2
    events, while recovering 81.6\% of the MODA expert-consensus events
    on the common excerpts. Despite stage-blind annotation, 99.0\%
    of ESW events occur in N2/N3. Used to train a fixed dense detector on a frozen SleepFM encoder, ESW reaches an event-level $F_1$ of $0.680$ on 166 unseen
    MODA subjects, compared with $0.635$ for E2 ($\Delta F_1=+0.045$, paired-bootstrap 95\% CI $[0.024,0.065]$).
    ESW is thus useful supervision for cross-dataset detection, and closer annotation-level agreement with the MODA consensus did not translate into better transfer performance in this experiment. The annotations are released under the MASS data-use terms (https://doi.org/10.5281/zenodo.23894975)


\end{abstract}

\section{Introduction}
Sleep spindles are transient 10--16\,Hz thalamocortical oscillations lasting
0.5 to 3\,s and are a hallmark of non-rapid-eye-movement (NREM) sleep
microstructure (stages N2 and N3)~\cite{iber2007aasm,fernandez2020spindles}.
They are described by five main characteristics (frequency, amplitude,
density, duration and topography), which differ between stage N2 (denser,
faster, centro-parietal spindles) and stage N3 (sparser, slower, more
frontal)~\cite{fernandez2020spindles}. For a given subject, those
characteristics reflect thalamocortical integrity, gray-matter thickness and
white-matter
myelination, and correlate with brain maturation, age and intelligence, as
well as with disorders leading to cognitive impairment or psychiatric
conditions such as dementia or schizophrenia~\cite{fernandez2020spindles}.
Characterizing how they vary with maturation and disease could, in the
long run, serve diagnostic purposes.

Manual annotation of spindles is too tedious for routine clinical
practice, and both the inclusion of events and their boundaries vary
between scorers~\cite{fernandez2020spindles,wendt2015reliability}. The
need for reliable automatic detection has therefore long been recognized,
from feature-based detectors~\cite{pallanca2018,lacourse2019a7} to recent
deep-learning models~\cite{kaulen2022sumo,guo2025blast}; yet in the MODA
benchmark, only two of the detectors tested reached the agreement of a
human expert~\cite{lacourse2020moda}. One reason is that the public
references cover the main spindle characteristics only partially. DREAMS
\cite{devuystdreams} was scored without stage
restriction, but on eight 30-min single-channel excerpts by two experts
whose agreement is below 0.5 (by-sample $F_1\approx0.42$), only one of
whom scored all excerpts. MASS-SS2~\cite{oreilly2014mass} provides 19
whole nights, but its two scorers annotated N2 only, on C3 only
\cite{oreilly2015benchmark}. MODA built a consensus of 5,342 spindles from
180 MASS subjects, but on 115-s blocks of artifact-free N2 displayed as 25-s
epochs of C3~\cite{lacourse2020moda}.
Hence topography cannot be measured on a single derivation; frequency, which
differs between N2 and N3, is sampled at one site and one stage; density
cannot be followed across a night or across stages on excerpts; and duration,
the least reproducible property between scorers~\cite{wendt2015reliability},
is shaped by filtering and consensus averaging. The authors of MODA mentioned
the need for a higher-validity reference that would require a complete
montage of channels, displaying a whole night per subject, and, since their
gold standard focuses on spindles in stage 2 and the C3 channel, no prior
selection of sleep stages~\cite{lacourse2020moda}.
Detecting the duration of a spindle, with a clear beginning and end, is
one of the most difficult tasks for scorers and algorithms alike, and at
the same time one of the most critical for describing the physiology:
duration, density and frequency are all defined on delimited events.
Epoch- or window-level classifiers only say whether a segment contains a
spindle; dense detectors instead predict a label for every sample, so
that event boundaries are learned from, and evaluated against, the
annotated onsets and offsets~\cite{tapia2020red,kaulen2022sumo}. On MODA,
the dense U-Net SUMO reaches an event $F_1$ of 0.82 at a temporal overlap
of 0.2, above the A7 detector (0.73) and the average individual expert
(0.72)~\cite{kaulen2022sumo}, and BLAST extends dense detection across
clinical settings~\cite{guo2025blast}. Among the dense architectures evaluated
in~\cite{bollengier2026icassp}, a detector combining a frozen
SleepFM encoder~\cite{thapa2026sleepfm} with a local
time--frequency branch and a dense decoder achieved an
event-level $F_1$ of $0.8080 \pm 0.0040$ at IoU 0.2 on MODA,
across six folds following the protocol
of~\cite{kaulen2022sumo}. This performance is comparable to
state-of-the-art detectors. We retain this architecture
unchanged, varying only the source of training annotations.
Predictions are evaluated at the event level using one-to-one
matching at a temporal IoU threshold of 0.2, following MODA
\cite{lacourse2020moda}. Unlike sample-level evaluation,
this protocol explicitly assesses individual event recovery
and temporal overlap.

Existing public references do not jointly provide stage-agnostic, full-montage,
whole-night expert spindle annotations. We fill this gap with
what we call \emph{spindles in the wild}: an expert annotation of complete
overnight recordings, across all stages, on the full EEG montage, blind to
the hypnogram and to prior annotations, with an explicit onset and offset for
every event; that is to say the very conditions that the authors of MODA identified as
missing from their own gold standard~\cite{lacourse2020moda}, and that
DREAMS had explored at small scale. MASS-SS2 is the natural test bed,
since a new annotation of its 19 nights can be compared, event by event,
with E1 and E2 and, on the 14 subjects shared with MODA, with the expert
consensus. Expert Spindles in the Wild (ESW) is, to our knowledge, the
largest such annotation to date ($\approx$150\,h of continuous EEG,
    $\approx$30k events; about 37 times the scored duration of DREAMS and six
times that of MODA). 
We contribute (i) the ESW annotation set; (ii) its characterization
against E1, E2 and the MODA consensus, separating event detection from
temporal delimitation, with a stage stratification of all events; and
(iii) a controlled transfer experiment showing that the same dense
detector generalizes better to unseen MODA subjects when supervised by
ESW than by E1, E2 or their union.

\section{Materials and methods}

The MASS dataset is an open archive of 200 whole-night
polysomnograms recorded at 256\,Hz, with EEG montages
comprising 4--20 channels, alongside EOG, EMG, ECG and
respiratory signals~\cite{oreilly2014mass}. Its SS2 subset
comprises 19 complete overnight recordings from 19 healthy
young adults, sleep-staged in 20-s epochs according to
standard scoring rules, totaling approximately 150\,h
of recordings.

Two historical spindle annotation sets are available, both
restricted to stage N2 and the C3 derivation
\cite{oreilly2015benchmark}. E1 followed standard scoring
rules on all 19 nights, yielding 10,017 events on the
15 nights shared with E2. In contrast, E2 annotated
15 nights using both broadband EEG and a sigma-filtered
(11--17\,Hz) signal, without imposing a minimum event
duration, yielding 22,254 events.

MODA scorers saw a single derivation (C3-A2 or C3-LE), one 25-s epoch at a
time, within 115-s blocks pre-selected from artifact-free N2, without the
possibility to move back and forth across epochs~\cite{lacourse2020moda}.
ESW lifts each of these restrictions. A senior board-certified sleep
neurophysiologist reviewed the 19 complete SS2 recordings from lights-off
to lights-on in a standard PSG viewer using 30-s epochs, with the signal
filtered between 0.3 and 35\,Hz and all EEG derivations displayed
and free navigation across the night. The hypnogram and the E1/E2 marks
were hidden, and sleep stage was not an inclusion criterion: the
instruction was to mark every event with spindle morphology wherever it
occurred, and to place its onset and offset explicitly rather than flag
its presence. Stage labels were attached a posteriori, without influence
on the scoring.

ESW contains 29,819 marks of positive duration; the 15 subjects shared
with E1 and E2 carry 27,013 ESW, 10,017 E1 and 22,254 E2 marks, E1 and E2
having been scored in N2 only, on C3~\cite{oreilly2015benchmark}. MODA
overlaps 14 SS2 subjects (41 blocks, 404 consensus events); requiring E2
as well leaves 12 subjects, 35 blocks and 370 events.


Agreement is measured with the MODA protocol~\cite{lacourse2020moda}.
Events are intervals compared by intersection over union (IoU); a
one-to-one matching is built greedily by descending IoU, yielding $M$
matched pairs, the pooled event $F_1=2M/(N_A+N_B)$ and the directional
recovery fractions $M/N_A$ and $M/N_B$. The primary threshold is
IoU${}=0.2$, the value at which MODA's own evaluation is defined; curves
over the full 0-1 range show how agreement depends on boundary
strictness. Against MODA, individual experts are evaluated as in
\cite{lacourse2020moda}, on the blocks they viewed and against the
leave-one-out consensus of the other scorers, whereas ESW, E1 and E2 are
evaluated against the full consensus on the common blocks; masks and
references differ, so the two sets of curves are shown side by side, not
as a ranking. 
Boundary conventions are summarized by the median onset, offset, and duration differences relative to the MODA expert consensus for reference events matched at IoU${}\geq0.2$ by all five sources (ESW, E1, E2, the MODA researcher consensus, and the MODA non-expert consensus), excluding events clipped in any source. These statistics describe temporal delimitation conditional on shared detection, rather than detection itself.
If ESW carries information that N2-restricted, single-channel annotations
lack, a detector trained on it should generalize better to an independent
reference. We test this with a dense detector adapted
from~\cite{bollengier2026icassp}: a frozen pre-trained SleepFM EEG
encoder~\cite{thapa2026sleepfm} provides temporal context, through
residual cross-attention, to a local time--frequency branch preserving
spindle-relevant spectral and intra-token temporal information, followed
by token-to-sample reconstruction and temporal refinement into
sample-level predictions. Seven supervision sets are compared-E1, E2,
ESW and their four unions-with identical architecture, hyper-parameters
and subject-level split of the 15 common SS2 nights; each model's decision
threshold is calibrated on its own validation subjects. Transfer is
evaluated zero-shot, without fine-tuning, on the 166 MODA subjects outside
SS2 (4,938 consensus events), by pooled event precision, recall and
micro-$F_1$ at IoU 0.2, with IoU 0.3 and 0.5 as sensitivity analyses.

The primary contrast is ESW versus E2 at IoU 0.2: E2, scored on a
sigma-filtered trace without minimum duration~\cite{oreilly2015benchmark},
is by protocol the most inclusive existing annotation of SS2 and hence the
most demanding comparator. A paired subject bootstrap (20,000 resamples of
the 166 subjects, TP/FP/FN re-pooled before computing the micro-$F_1$
difference) gives pointwise percentile 95\% intervals; permutation
$p$-values over the 21 model pairs are Holm-adjusted. Inference is
conditional on the fitted models and does not account for retraining
variability.


\begin{figure}[t]
    \centering
    \setlength{\panelwidth}{.49\linewidth}
    \panel{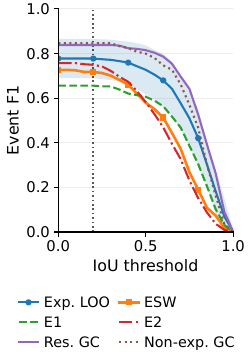}{a}{Agreement
    with the MODA consensus.}\hfill
    \panel{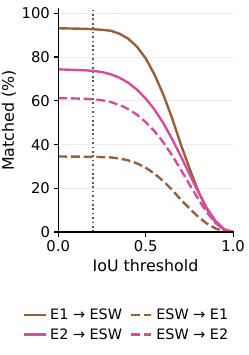}{b}{Directional
    recovery, 15 common nights.}
    \caption{(a)~Event $F_1$ against the MODA expert consensus vs.\ IoU
        threshold on the 35 common blocks (12 subjects, 370 events): ESW, E1, E2
        against the full consensus; mean $\pm$ SD of the 23 individual experts,
        each leave-one-out on the blocks it viewed (masks and references differ:
        not a ranking on identical data). (b)~Fraction of E1 and E2 events
        recovered by ESW (solid) and of ESW events recovered by E1 and E2
    (dashed) on the 15 common nights.}
    \label{fig:overview}
\end{figure}

\section{Results}

\textbf{ESW covers the historical annotations and extends them.} On the
15 common nights, ESW matches 92.7\% of E1 events (9,286/10,017) and
73.7\% of E2 events (16,392/22,254), whereas E1 and E2 recover only 34.4\%
and 60.7\% of ESW (Fig.~\ref{fig:overview}b). Pairwise $F_1$ at IoU 0.2 is
.592 for E1--E2, .502 for E1--ESW and .665 for E2--ESW: ESW agrees better
with E2 than the two historical scorers agree with each other. ESW is
broader but not a superset---731 E1 and 5,862 E2 events have no ESW
match---and its net excess over E2 (4,759 events, $+21.4\%$) is a count of
additional candidates, not of validated spindles.

\textbf{ESW recovers a similar proportion of MODA consensus events to E2, while proposing more unmatched candidates.} On the 35 common blocks, ESW recovers 302/370
consensus events (81.6\%), as many as E2 (82.4\%) and far more than E1
(50.0\%), with more unmatched candidates and hence a lower precision
(Table~\ref{tab:human}); on all 14 overlapping subjects, 319/404 (79.0\%,
$F_1$ .708). ESW lies within the band of the 23 individual MODA experts
(leave-one-out $F_1$ .777 $\pm$ .088; Fig.~\ref{fig:overview}a). The
higher scores of MODA's researcher and non-expert group consensuses (GC,
.838 and .846) reflect the aggregation of several scorers of the same C3
epochs, not a single scorer~\cite{lacourse2020moda}.
\begin{table}[t]
    \centering\small
    \setlength{\tabcolsep}{4pt}
    \caption{Agreement against the MODA expert consensus at IoU 0.2 (12
        subjects, 35 blocks, 370 consensus events). $N$: annotations in the
        blocks; TP: matched; U: unmatched; $P$: precision; $R$: recall.
    ~\cite{lacourse2020moda}.}
    \label{tab:human}
    \begin{tabular}{lrrrrrr}\toprule
        Source & $N$ & TP & U & $P$ & $R$ & $F_1$\\\midrule
        E1  & 194 & 185 &   9 & .954 & .500 & .656\\
        E2  & 444 & 305 & 139 & .687 & .824 & .749\\
        ESW & 472 & 302 & 170 & .640 & .816 & .717\\\midrule
        GC researchers & 322 & 290 &  32 & .901 & .784 & .838\\
        GC non-experts & 353 & 306 &  47 & .867 & .827 & .846\\\bottomrule
    \end{tabular}
\end{table}

\textbf{Scored blind to stage, ESW reproduces the stage specificity of
spindles.} Assigned a posteriori to the hypnogram, the 29,819 events fall
82.3\% in N2 (24,540), 16.7\% in N3 (4,980), 0.6\% in N1 (180), 0.3\% in
REM (78) and 0.1\% in W (37), with four unassigned: 99.0\% in N2/N3.
Densities per stage-hour are 338 in N2, 204 in N3, 21.9 in N1, 2.9 in REM
and 2.4 in W. 



\textbf{Boundary conventions contribute to disagreement at stricter IoU thresholds.} The three curves of Fig.~\ref{fig:overview}a diverge as the
threshold tightens, in the order of the annotators' event durations: on
the 35 common blocks, mean durations are 0.79\,s for E1, 1.07\,s for ESW
and 1.16\,s for E2, against 0.87\,s for the consensus. On the 149
consensus events matched by all sources at IoU${}\geq0.2$, median duration
differences to the consensus are $-142$\,ms for E1, $+116$\,ms for ESW and
$+339$\,ms for E2; ESW's onset is 104\,ms earlier in median while its
offset is unchanged ($-1$\,ms), so its wider interval comes from an
earlier onset rather than a symmetric extension. Consistently, the
fraction of these events still matched at IoU 0.7 is 71.1\% for E1, 61.1\%
for ESW and 49.7\% for E2. Duration mismatch bounds the attainable IoU
($\mathrm{IoU}\le\min(d_S,d_G)/\max(d_S,d_G)$), but the full $F_1$ also
counts misses and unmatched marks: at IoU 0.5, ESW matches 244 consensus
events against 171 for E1, yet scores .580 against .606 because it
proposes more candidates. The curves thus reflect distinct temporal
conventions, without indicating which is physiologically correct or how
much of the $F_1$ gap they explain.



\textbf{ESW is better supervision for transfer to MODA than E1, E2 or
their union.} Trained on the same 15 nights with the same detector, the
models transfer zero-shot to the 166 MODA subjects outside SS2 (4,938
consensus events) with an event $F_1$ at IoU 0.2 of .680 for ESW, .635
for E2 and .328 for E1 (Table~\ref{tab:learning}). The ESW--E2 difference
is $+.045$ (paired subject bootstrap 95\% CI [.024, .065], Holm-adjusted
$p=.001$), with gains in both precision ($+3.8$ points) and recall
($+5.1$ points); it persists at IoU 0.3 ($+.046$, CI [.025, .066]) and
0.5 ($+.040$, CI [.019, .060]), and subject-level $F_1$ favors ESW in 103
subjects against 55 for E2. The E1-trained model inherits the
conservativeness of its protocol (precision .972, recall .197). Pooling
the historical annotations (E1$\cup$E2, .653) does not reach ESW alone,
and adding E1 or E2 to ESW changes $F_1$ by at most $+.004$ (ESW$\cup$E1,
CI [$-$.001, .009], adjusted $p=.55$): Adding historical annotations did not provide a statistically significant improvement over ESW alone at the primary IoU threshold.

\begin{table}[t]
    \centering\small
    \setlength{\tabcolsep}{3pt}
    \caption{Zero-shot transfer to the 166 MODA subjects outside SS2 (4,938
        consensus events): event $F_1$ at IoU 0.2 with subject-bootstrap 95\% CI
    (marginal, not for differences), and at IoU 0.5.}
    \label{tab:learning}
    \begin{tabular}{lrcr}\toprule
        Training labels & $F_1$@.2 & 95\% CI & $F_1$@.5\\\midrule
        E1 & .328 & [.274, .379] & .307\\
        E2 & .635 & [.609, .659] & .561\\
        ESW & \textbf{.680} & [.657, .700] & \textbf{.601}\\
        E1$\cup$E2 & .653 & [.629, .674] & .570\\
        ESW$\cup$E1 & .684 & [.660, .705] & .609\\
        ESW$\cup$E2 & .680 & [.659, .700] & .567\\
        ESW$\cup$E1$\cup$E2 & .683 & [.661, .703] & .591\\\bottomrule
    \end{tabular}
\end{table}


\section{Discussion and conclusion}

Three findings are distinct. ESW broadens coverage: it recovers most of E1
and E2 and as many MODA consensus events as E2, while adding candidates
that no reference validates. Its boundaries differ systematically from
E1 and E2, which may contribute to the loss of agreement at strict IoU
without establishing which convention is more accurate. Finally, ESW
provides better supervision for the tested detector than E1, E2 or their
union. Importantly, this third finding does not follow directly from
the first two: higher training utility coexists here with lower
agreement with the MODA consensus than E2.

The improved performance of ESW supervision may arise from three
plausible mechanisms. First, its stage-agnostic annotation protocol may
provide a broader representation of spindle and non-spindle EEG
activity, potentially helping the detector distinguish spindles from
other oscillatory patterns, including alpha and REM-related activity.
Second, ESW's intermediate onset and offset conventions may be more
compatible with the MODA consensus. Third, the additional spindle
candidates identified by ESW do not appear to impair transfer, as the
ESW-trained model achieves both higher precision and recall than the
E2-trained model. The relative contributions of these mechanisms
remain to be established. The E1-trained model, meanwhile, retains
the conservativeness of its annotation protocol on unseen data.

MODA and ESW are complementary rather than competing: MODA pooled
17,367 technologist, 7,613 researcher and 63,935 non-expert marks
over about 24\,h of N2 blocks~\cite{lacourse2020moda}, establishing
a consensus across multiple scorers. In contrast, ESW reflects a
single expert's assessment of $\approx$150\,h of continuous EEG,
providing broader temporal and sleep-stage coverage. The stage distribution supports the physiological consistency of the
annotation protocol: despite being blind to the hypnogram, the expert
placed 99\% of events in N2/N3, with event densities approximately two
orders of magnitude higher than in wake and REM. The remaining 295 events
outside N2/N3 are consistent with reports of sigma activity at sleep
onset, low-density spindles during REM sleep in intracranial recordings,
and mixed sleep states~\cite{fernandez2020spindles,andrillon2011,%
bandarabadi2020}, but they have not been independently adjudicated and
their physiological significance remains to be established.

\textbf{Limitations.}  ESW has one annotator and no intra-scorer
reliability estimate yet. The boundary analysis is conditional on shared
detection, and its 149 events are longer than average. The gain cannot
yet be attributed to coverage, boundaries or channel context: MODA
contains N2 only and cannot reveal what ESW adds outside N2, and the
ablations that would separate these factors (ESW restricted to N2, ESW
with E1-like boundaries, stage-stratified evaluation on whole nights) are
only possible because ESW covers the whole night. Inference is
conditional on the fitted models, and both source and target originate
in MASS, so this is subject-disjoint transfer, not validation on an
independent acquisition domain.

ESW is, to our knowledge, the largest stage-agnostic whole-night expert
annotation of sleep spindles, validated against two independent
references and demonstrably useful as supervision for a foundation model, though not yet as a new physiological gold standard. As the annotations are released, the additional
events can be adjudicated and the mechanisms of the gain isolated by
controlled ablation.

\clearpage
\section{Compliance with ethical standards}
MASS recordings and associated annotations were used in accordance with the
archive's data-use agreement. The study was conducted under protocol
CER-2025-Navarro-PREDSomADICE-00578, which received a favorable opinion from
the Research Ethics Committee of Sorbonne Universit\'e.

\section{Acknowledgments and conflicts of interest}
We are grateful to the Fondation AP-HP, which funded this work through
the DREAM project. The project was initiated in the Medical Artificial
Intelligence group of the Idiap Research Institute with the participation of
Vrije Universiteit Brussel; we thank Idiap for computing resources and
for facilitating access to the MASS recordings. The authors declare no
competing interests. The authors used ChatGPT (OpenAI) and Claude
(Anthropic) to assist with drafting and editing the text and images; all
annotations, analyses, results and conclusions are the authors' own, and
the authors take full responsibility for the manuscript.


\bibliographystyle{IEEEbib}
\bibliography{references}

@article{fernandez2020spindles,
    author  = {Fernandez, L. M. J. and L{\"u}thi, A.},
    title   = {Sleep Spindles: Mechanisms and Functions},
    journal = {Physiological Reviews},
    year    = {2020},
    volume  = {100},
    number  = {2},
    pages   = {805--868},
    doi     = {10.1152/physrev.00042.2018}
}

@article{oreilly2014mass,
    author  = {O'Reilly, C. and others},
    title   = {Montreal Archive of Sleep Studies: An Open-Access Resource for Instrument Benchmarking and Exploratory Research},
    journal = {Journal of Sleep Research},
    year    = {2014},
    volume  = {23},
    number  = {6},
    pages   = {628--635},
    doi     = {10.1111/jsr.12169}
}

@article{lacourse2020moda,
    author  = {Lacourse, K. and others},
    title   = {Massive Online Data Annotation, Crowdsourcing to Generate High Quality Sleep Spindle Annotations from {EEG} Data},
    journal = {Scientific Data},
    year    = {2020},
    volume  = {7},
    pages   = {190},
    doi     = {10.1038/s41597-020-0533-4}
}

@article{thapa2026sleepfm,
    author  = {Thapa, R. and others},
    title   = {A Multimodal Sleep Foundation Model for Disease Prediction},
    journal = {Nature Medicine},
    year    = {2026},
    volume  = {32},
    number  = {2},
    pages   = {752--762},
    doi     = {10.1038/s41591-025-04133-4}
}

@article{guo2025blast,
    author  = {Guo, C. and others},
    title   = {{BLAST}: Towards Robust Detection of Sleep Spindles Across Clinical Settings},
    journal = {Neurocomputing},
    year    = {2025},
    volume  = {653},
    pages   = {131107},
    doi     = {10.1016/j.neucom.2025.131107}
}

@article{kaulen2022sumo,
    author  = {Kaulen, L. and others},
    title   = {Advanced Sleep Spindle Identification with Neural Networks},
    journal = {Scientific Reports},
    year    = {2022},
    volume  = {12},
    number  = {1},
    pages   = {7686},
    doi     = {10.1038/s41598-022-11210-y}
}

@article{oreilly2015benchmark,
    author  = {O'Reilly, C. and Nielsen, T.},
    title   = {Automatic Sleep Spindle Detection: Benchmarking with Fine Temporal Resolution Using Open Science Tools},
    journal = {Frontiers in Human Neuroscience},
    volume  = {9},
    pages   = {353},
    year    = {2015},
    doi     = {10.3389/fnhum.2015.00353}
}

@misc{devuystdreams,
    author       = {Devuyst, S.},
    title        = {The {DREAMS} Databases and Assessment Algorithm},
    howpublished = {Zenodo},
    year         = {2019},
    doi          = {10.5281/zenodo.2650141},
    note         = {Dataset. Accessed 23 Oct. 2026}
}

@article{wendt2015reliability,
    author  = {Wendt, S. L. and others},
    title   = {Inter-Expert and Intra-Expert Reliability in Sleep Spindle Scoring},
    journal = {Clinical Neurophysiology},
    volume  = {126},
    number  = {8},
    pages   = {1548--1556},
    year    = {2015},
    doi     = {10.1016/j.clinph.2014.10.158}
}

@unpublished{bollengier2026icassp,
    author = {Bollengier, A. and D{\'i}az Berenguer, A. and Pallanca, O. and Anjos, A. and Sahli, H.},
    title  = {Adapting Sleep Foundation Models for Dense Sleep Spindle Localization},
    year   = {2026},
    note   = {Submitted to IEEE ICASSP 2027, under review}
}

@book{iber2007aasm,
    author    = {Iber, C. and Ancoli-Israel, S. and Chesson, A. and Quan, S. F.},
    title     = {The {AASM} Manual for the Scoring of Sleep and Associated Events: Rules, Terminology and Technical Specifications},
    publisher = {American Academy of Sleep Medicine},
    address   = {Westchester, IL},
    year      = {2007}
}

@article{lacourse2019a7,
    author  = {Lacourse, K. and Delfrate, J. and Beaudry, J. and Peppard, P. and Warby, S. C.},
    title   = {A Sleep Spindle Detection Algorithm That Emulates Human Expert Spindle Scoring},
    journal = {Journal of Neuroscience Methods},
    year    = {2019},
    volume  = {316},
    pages   = {3--11},
    doi     = {10.1016/j.jneumeth.2018.08.014}
}

@inproceedings{tapia2020red,
    author    = {Tapia, N. I. and Est{\'e}vez, P. A.},
    title     = {{RED}: Deep Recurrent Neural Networks for Sleep {EEG} Event Detection},
    booktitle = {2020 International Joint Conference on Neural Networks (IJCNN)},
    year      = {2020},
    pages     = {1--8},
    doi       = {10.1109/IJCNN48605.2020.9207719}
}

@inproceedings{pallanca2018,
    author    = {Pallanca, O. and Khalife, S. and Read, J.},
    title     = {Detection of Sleep Spindles in {NREM} 2 Sleep Stages: Preliminary Study \& Benchmarking of Algorithms},
    booktitle = {2018 IEEE International Conference on Bioinformatics and Biomedicine (BIBM)},
    year      = {2018},
    pages     = {2652--2655},
    doi       = {10.1109/BIBM.2018.8621305}
}

@article{andrillon2011,
    author  = {Andrillon, T. and Nir, Y. and Staba, R. J. and Ferrarelli, F. and Cirelli, C. and Tononi, G. and Fried, I.},
    title   = {Sleep Spindles in Humans: Insights from Intracranial {EEG} and Unit Recordings},
    journal = {Journal of Neuroscience},
    year    = {2011},
    volume  = {31},
    number  = {49},
    pages   = {17821--17834},
    doi     = {10.1523/JNEUROSCI.2604-11.2011}
}

@article{bandarabadi2020,
    author  = {Bandarabadi, M. and Guti{\'e}rrez Herrera, C. and Gent, T. C. and Bassetti, C. and Schindler, K. and Adamantidis, A. R.},
    title   = {A Role for Spindles in the Onset of Rapid Eye Movement Sleep},
    journal = {Nature Communications},
    year    = {2020},
    volume  = {11},
    pages   = {5247},
    doi     = {10.1038/s41467-020-19076-2}
}
\end{document}